\documentclass{article}

\usepackage{ijcai13}
\usepackage{times,graphicx}

\title{Flavor Pairing in Medieval European Cuisine: \\A Study in Cooking with Dirty Data}
\author{Kush R. Varshney,$^1$ Lav R. Varshney,$^1$ Jun Wang,$^1$ and Daniel Myers$^2$\\
$^1$IBM Thomas J. Watson Research Center\\
1101 Kitchawan Road, Yorktown Heights, NY, USA\\
\{krvarshn,lrvarshn,wangjun\}@us.ibm.com\\
$^2$Medieval Cookery\\
dmyers@medievalcookery.com}

\begin{document}

\maketitle

\begin{abstract}
An important part of cooking with computers is using statistical methods to create new, flavorful ingredient combinations.  The flavor pairing hypothesis states that culinary ingredients with common chemical flavor components combine well to produce pleasant dishes.  It has been recently shown that this design principle is a basis for modern Western cuisine and is reversed for Asian cuisine.  

Such data-driven analysis compares the chemistry of ingredients to ingredient sets found in recipes.  However, analytics-based generation of novel flavor profiles can only be as good as the underlying chemical and recipe data.  Incomplete, inaccurate, and irrelevant data may degrade flavor pairing inferences.  Chemical data on flavor compounds is incomplete due to the nature of the experiments that must be conducted to obtain it.  Recipe data may have issues due to text parsing errors, imprecision in textual descriptions of ingredients, and the fact that the same ingredient may be known by different names in different recipes.  Moreover, the process of matching ingredients in chemical data and recipe data may be fraught with mistakes.  Much of the `dirtiness' of the data cannot be cleansed even with manual curation.  

In this work, we collect a new data set of recipes from Medieval Europe before the Columbian Exchange and investigate the flavor pairing hypothesis historically.  To investigate the role of data incompleteness and error as part of this hypothesis testing, we use two separate chemical compound data sets with different levels of cleanliness.  Notably, the different data sets give conflicting conclusions about the flavor pairing hypothesis in Medieval Europe.  As a contribution towards social science, we obtain inferences about the evolution of culinary arts when many new ingredients are suddenly made available.
\end{abstract}

``Computational creativity is a subfield of artificial intelligence research \ldots where we build and work with computational systems that create art[i]facts and ideas. These systems are usually, but not exclusively, applied in domains historically associated with creative people, such as mathematics and science, poetry and story telling, musical composition and performance, video game, architectural, industrial and graphic design, the visual, and even the culinary, arts'' \cite{ColtonW2012}.

\section{Introduction}
\label{sec:intro}

The cooking of food and human evolution are intertwined.  One could go so far as to argue that it is cooking that makes us human.  We are naturally drawn to foods high in fat and sugar because of the nourishment we received from such foods during our evolution in resource-poor environments.  However, we are also drawn to foods with complex layers of balanced flavors composed through the art and science of cooking.  It is these foods that we find delectable, delightful, and delicious.

Human flavor perception is very complicated, involving a variety of external sensory stimuli and internal states \cite{Shepherd2006,LimJ2012}.  Not only does it involve the five classical senses, but also sensing through the gut, and the emotional, memory-related, motivational, and linguistic aspects of food.  In addition to the basic tastes: sweet, sour, salty, bitter, and umami, the smell of foods is the key contributor to flavor perception, which is in turn a property of the chemical compounds contained in the ingredients \cite{Burdock2004}.  There are typically tens to hundreds of different flavor compounds per food ingredient \cite{AhnABB2011}.  

Other contributors to flavor perception among the classical senses are the temperature, texture, astringency, and creaminess of the food; the color and shape of food; and the sound that the food makes.  The digestive system detects the autonomic and metabolic properties of the food.  Moreover, there are emotion, motivation, and craving circuits in the brain that influence flavor perception, which are in turn related to language, feeding, conscious flavor perception, and memory circuits.  Furthermore, effects beyond the food itself, including social and contextual ones, influence flavor perception \cite{KingMC2013}.

The key quality of foods and the one we focus on in this paper is the set of flavor compounds that they contain, which is a union of the flavor compounds of the constituent ingredients.  Recent work has shown that olfactory pleasantness can be predicted based on the structure of flavor compounds \cite{KhanLFALHS2007,LapidHS2008,HaddadMRHS2010}.  We note that cooking ingredients together can influence flavor perception as proteins, fats and starches bind certain flavor compounds and certain compounds may evaporate away or be chemically changed, but this is a second-order effect which we do not study in this work \cite{Guichard2002}.

One of the main guiding principles in putting together a recipe or a dish is flavor pairing.  It is believed that ingredients that share many flavor compounds go well together.  This flavor pairing hypothesis arose when the chef Heston Blumenthal found caviar and white chocolate to go well together, and investigated the basic chemical reason for why this is a good pairing \cite{Blumenthal2008}.  The flavor pairing hypothesis has been scientifically studied for several modern cuisines and found to hold strongly for Western cuisine, but to be almost opposite in East Asian cuisine \cite{AhnABB2011}.

In interviews that we conducted with professional chef instructors from the Institute of Culinary Education, we found that cutting-edge chefs today do truly think in terms of pairs or triplets of ingredients when coming up with new recipes.  While they are not explicit about the chemistry of flavor compounds, they do draw on their mental databases of flavors and combinations to pair ingredients. They also draw upon other types of similarities between ingredients, such as being grown in the same season of the year or in the same region of the world.

One way in which we can use computers in cooking is as an aid in coming up with new ingredient pairings that may lie outside of the chef's mental repository.  Having a machine system generate completely new flavor combinations that have never existed before is a culinary application of computational creativity \cite{BhattacharjyaVPC2012,MorrisBBV2012,VeeramachaneniVO2012,VarshneyPVSC2013}.  Such a culinary computational creativity system or synthetic gastronomist depends critically on its data repository (just like human culinary creators).

The quality of data (or lack thereof) is a prevalent issue in almost all fields of analytics \cite{KimCHKL2003,DeVeauxH2005,WangSML2012}.  Here we list several aspects of this issue as they specifically pertain to the culinary domain approached with computers.  One issue is resolving the names of ingredients that refer to the same entity, e.g.~`emmental,' `emmenthal' and `emmenthaler'; `bow tie pasta,' `bowtie pasta' and `farfalle'; `cilantro,' `coriander' and `dhania'; `New Mexico red chile' and `red New Mexico chile'; and `achiote,' `annatto' and `annatto seed.'  In parsing and analyzing semi-structured recipe text, there can be issues in determining what the actual ingredient is, e.g. in `my low-carb catsup,' and other aspects of data dirtiness arising from imperfect text analytics.

Moreover, the analytical chemistry experiments conducted to determine the flavor compounds present in a food ingredient are far from perfect and repositories far from complete.  Thermal desorption--gas chromatography--mass spectrometry is the typical technique used to identify and sometimes quantify volatile flavor compounds in foods and beverages, but also in many forensics, monitoring, and other applications.  Experiments have not been conducted and verified for every possible food ingredient, which presents a missing data limitation.  Also, many compounds occur in trace amounts in foods; this contributes to false alarm and missed detection errors in chemical analysis.  Moreover, matching food ingredients in recipes to food ingredients that have been chemically analyzed is another process that can introduce error and incompleteness.  Thus overall, recipe, ingredient, and flavor compound data is plagued by various kinds of data dirtiness.

In order to cook with computers, specifically by having the computer use the flavor compound pairing hypothesis to create new sets of ingredients in recipes, we must understand the effect of data dirtiness.  Towards that end, in this paper we obtain two independent data sets of flavor compound data having different levels of completeness and accuracy, and compile a completely new corpus of recipes from the Late Middle Ages that has not been analyzed before.  We test the flavor pairing hypothesis twice using this corpus: once with each flavor compound data set.  With one data set we see a positive confirmation of the flavor pairing hypothesis that is off the charts, but see an opposite result using the other data set.  These conflicting conclusions arising from differences in data quality are quite notable and must be scrutinized in the process of cooking with computers.

A second contribution of our work beyond understanding dirty data when cooking with computers is a statement about the evolution of cuisine \cite{KinouchiDHZR2008}.  The medieval period represents an age just before the set of possible ingredients for European cooking increased dramatically.  In the age of discovery that followed, European exploration of the Americas initiated the Columbian Exchange \cite{Crosby1972}, the transfer of animals, plants, bacteria, viruses, and culture between the continents.  This transfer represents the most significant global event in terms of agriculture and cooking because it injected many now-common ingredients such as tomatoes and potatoes into Eurasian and African cuisine.  When compared to modern Western cuisine using the same flavor compound data set, we find that the level of flavor pairing is approximately the same as in medieval cuisine, but that the ingredients available do not lend themselves as much to chemical pairing.  

The remainder of this paper is organized as follows.  In Section~\ref{sec:pairing}, we discuss the flavor pairing hypothesis in greater depth and describe the statistical methodology for testing it and the data required for such evaluation.  Then we discuss the medieval period in Europe and the Columbian Exchange in Section~\ref{sec:medi}.  Section~\ref{sec:empirical} is devoted to empirical methodology and results: testing flavor pairing on medieval recipes using two different flavor compound data sets.  The main result in this section is that data quality issues can yield conflicting inferences.  We provide discussion and conclude in Section~\ref{sec:conclusion}.

\section{Flavor Pairing Hypothesis Testing}
\label{sec:pairing}

As mentioned in the introduction, flavor pairing is a key concept in culinary arts that can be examined scientifically by investigating the chemical flavor compounds that are components of food ingredients.  In this section, we discuss the different aspects of testing and quantifying the extent to which ingredients with common flavor compounds go well together.

\subsection{Chemical Components}
\label{sec:pairing:chem}

A strong determinant of the flavor of foods is the aromatic compounds that reach the olfactory system, either through the nose or through retro-olfaction.  Humans are adept at detecting even trace amounts of these compounds and they have a great effect on hedonic perception.  Chemically, flavor compounds come from groups such as acetals, acids, alcohols, aldehydes, esters, furans, hydrocarbons, ketones, lactones, and phenols.

The processes by which one can determine the flavor components of a food are a branch of analytical chemistry.  The typical experiment involves heating the food to release the flavor compounds into a vapor.  This gas is then passed through a gas chromatograph, which separates different types of molecules of the gas based on the time it takes them to travel through a capillary.  The separated gas molecules are then analyzed using a mass spectrometer, which ionizes the molecules and measures mass-to-charge ratios to determine the elemental composition.

In such a manner, the flavor compounds that are present in a food dish, or more typically food ingredient, are identified.  However, when many different flavor compounds are present in small quantities, the experiments generally have some false alarms and some missed detections.  In any case, the result of the analytical chemistry is a list of contained compounds for the food ingredient under consideration.  In our work, we consider two flavor compound databases: the Volatile Compounds in Food 14.1 database (VCF) and Fenaroli's Handbook of Flavor Ingredients as processed and released in \cite{AhnABB2011}.

\subsection{Recipe Collections}
\label{sec:pairing:rec}

Recipes in cookbooks represent the culinary best practices of a culture.  As such, when taking a data-driven approach to understanding pairing, they also represent sets of ingredients that are flavorful together.  Modern recipes list ingredients, but also quantities and instructions for preparation.  In the medieval period, recipes were primarily only the former.  For the purposes of analyzing pairing, it is only the set of ingredients that is of importance.  If the flavor pairing hypothesis holds, then sets of ingredients in actual recipes should have, on average, more shared flavor compounds than any random set of ingredients.  It is such a test that is proposed in \cite{AhnABB2011} and that we conduct in this work.

We note that rather than using a large-scale statistical methodology with the foundational assumption that recipes in cookbooks represent the distillation of what people like and dislike, another approach is experimental.  In a very small-scale sensory testing experiment, twenty-one food pairings involving pear, tomato, potato, chocolate, beef, cauliflower, and anise were made into pur\'ees and tested using human flavor perception experiments.  Combinations with strong flavor pairing according to the VCF database were not necessarily the best rated by undergraduate test subjects \cite{KortNID2010}.

In this paper, we compile a corpus of recipes from Medieval Europe and analyze the flavor pairing hypothesis within this new collection.  There are some notable problems with drawing conclusions based on recipes in cookbooks from the time if one's goal is to understand daily life in medieval times. Most notably, the cookbooks are of wealthy landowners, and thus do not necessarily reflect the diet of the poor or middle class. Also, the connection between recipes and what was actually cooked is open to question. However, our goal is to understand the most flavorful foods that were being concocted in that time and place, and recipes are an excellent source for that purpose. 

\subsection{Statistical Methodology}
\label{sec:pairing:math}

As described in \cite{AhnABB2011}, the primary calculation to understand the flavor pairing hypothesis is to compute the average number of shared flavor compounds among the ingredients in a recipe $R$.  Let $R$ be a set of $n_R$ different ingredients.  Then the average number of shared compounds is:
\begin{equation}
\label{eq:sharedcompounds}
	N_s(R) = \frac{2}{n_R(n_R-1)} \sum_{i,j\in R,i\neq j} |C_i \cap C_j |,
\end{equation}
where $C_i$ is the set of flavor compounds in ingredient $i$ and $C_j$ is the set of flavor compounds in ingredient $j$.  The mean of $N_s(R)$ across the corpus of recipes, which we denote $\bar{N}_s$, represents the degree to which flavor pairing exists overall.  

Then in order to understand whether $\bar{N}_s$ is indicative of ingredients with high compound sharing also often appearing together in recipes (and thus implicitly tasting good together), we must compare $\bar{N}_s$ for the recipe corpus under consideration to a null hypothesis, specifically the value of $\bar{N}_s$ for randomly generated sets of ingredients from the same overall universe of ingredients and probability distribution.  Denoting the average sharing for the true corpus of recipes as $\bar{N}_s^{\rm{real}}$ and for a randomly generate corpus as $\bar{N}_s^{\rm{rand}}$, the difference
\begin{equation}
\label{eq:deltaNs}
	\Delta N_s = \bar{N}_s^{\rm{real}} - \bar{N}_s^{\rm{rand}}
\end{equation}
if positive indicates that flavor pairing is a strong influence in the real recipes under consideration, if close to zero indicates no relationship between flavor compounds and recipes, and if negative indicates that recipes tend to include ingredients that spcifically do not share flavor compounds.

Additionally, as discussed extensively in \cite{AhnABB2011}, it is possible to calculate how much an individual ingredient $i$ contributes to $\Delta N_s$ as follows:
\begin{equation}
\label{eq:contribution}
	\chi_i = \frac{1}{N_c}\sum_{R \ni i} N_s(R) - \left(\frac{2f_i}{N_c\bar{n}_R}\frac{\sum_{j\in c} f_j |C_i \cap C_j|}{\sum_{j\in c} f_j}\right),
\end{equation}
where $N_c$ is the number of recipes in the corpus, $f_i$ is the number of occurrences of ingredient $i$, and $\bar{n}_R$ is the average number of ingredients per recipe in the corpus.

\section{Medieval Times and the Columbian Exchange}
\label{sec:medi}

The medieval period in Europe, also known as the Middle Ages, is the time between the collapse of the Western Roman Empire and the beginning of the  Renaissance.  The exact dates are a bit hard to pin down, and strongly depend on what part of Europe is being considered. For example the fifteenth century is considered as the Renaissance in Italy, but is the Late Middle Ages in England. 

Cereal grains (barley, oats, and rye for the poor and wheat for the wealthy) were the main staples and were prepared as bread, porridge, gruel, and pasta.  The staples were supplemented by vegetables.  Meat was more expensive and eaten less, with pork and chicken being more prevalent than beef.  Fish was common, especially cod and herring, but also other saltwater and freshwater fish.  Wild game was common only among the nobility.

A misconception about that time period is that spices were used to cover the taste of spoiled meat. This myth has its origins in Victorian-era England and has no basis in fact. Such a practice would have been unfeasible in terms of health (it would have killed the people), economics (it would have been too expensive), and logistics (it would have required vast amounts of meat to be kept hanging for days).

The medieval period was an age prior to the exploration of the Americas.  Once the New World had been discovered, many new ingredients made their way to Eurasia and vice versa.  This transfer of foods, along with the transfer of diseases and culture is known as the Columbian Exchange after Christopher Columbus \cite{Crosby1972,NunnQ2010}.  Some key ingredients that were absent in the Old World before the Columbian Exchange include corn, potatoes, cassava, sweet potatoes, tomatoes, sunflower seeds, cacao beans, pineapples, peanuts, eggplants, tobacco, vanilla, and capsicum peppers (which are the ancestors of cayenne peppers, bell peppers, and jalape\~no peppers).  Crops such as tomatoes, cacao, and chili peppers are not themselves especially rich in calories, but complement existing foods by increasing vitamin intake and improving flavor. 
\begin{table}[t]
	\begin{center}
		\begin{tabular}{|p{4.4cm}|l|l|} \hline
			Book & Country & Date \\\hline\hline
			MS B.L. Royal 12.C.xii & England/ & 1340 \\
			& France & \\\hline
			Forme of Cury & England & 1390 \\\hline
			Ancient Cookery & England & 1425 \\\hline
			Liber cure cocorum & England & 1430 \\\hline
			Two Fifteenth-Century Cookery Books & England & 1450 \\\hline
			A Noble Boke off Cookry & England & 1468 \\\hline
			Thomas Awkbarow's Recipes [MS Harley 5401] & England & 15th c. \\\hline
			Gentyll manly Cokere [MS Pepys 1047] & England & ca. 1500 \\\hline
			A Proper newe Booke of Cokerye & England & 1550 \\\hline
			A Book of Cookrye & England & 1591 \\\hline
			The Good Housewife's Jewell & England & 1596 \\\hline
			Delights for Ladies & England & 1609 \\\hline
			A NEVV BOOKE of Cookerie & England & 1615 \\\hline\hline
			Enseignements & France & 1300 \\\hline
			Le Viandier de Taillevent & France & 1380 \\\hline
			Le Menagier de Paris & France & 1393 \\\hline
			Du fait de cuisine & France & 1420 \\\hline
			Le Recueil de Riom & France & 15th c. \\\hline
			Ouverture de Cuisine & France & 1604 \\\hline\hline
			Ein Buch von guter spise & Germany & 1345 \\\hline
			Das Kochbuch des Meisters Eberhard & Germany & 1450 \\\hline
			Das Kuchbuch der Sabina Welserin & Germany & 16th c. \\\hline\hline
			Libro di cucina / Libro per cuoco & Italy & 14th/15th c. \\\hline
			The Neapolitan recipe collection & Italy & 15th c. \\\hline
			Due Libri di Cucina - Libro B & Italy & 15th c. \\\hline			
		\end{tabular}
	\end{center}
	\caption{Medieval source texts.}
	\label{table:medieval_cookbooks}
\end{table}

Often New World foods have had an important effect on cuisine evolution: chili peppers led to spicy curries in India, paprika in Hungary, and spicy kimchee in Korea; tomatoes significantly altered the cuisine of Italy and other Mediterranean countries.  Thus it is interesting to examine cuisine from before the exchange to understand culinary evolution \cite{KinouchiDHZR2008}.

\section{Empirical Methodology and Results}
\label{sec:empirical}

In this section, we describe the steps we undertook to empirically study the flavor pairing hypothesis in Medieval Europe, from constructing the corpus of recipes all the way to conducting the analytics.  These are the same steps that need to be performed when cooking with a computer that suggests new food pairings based on cultural artifacts and chemistry.
\begin{table}[t]
	\begin{center}
		\begin{tabular}{|c|c|c|c|c|} \hline
			(a) & (b) & (c) & (d) & (e)\\\hline
			bean & venison & eel & mallard & frumenty\\
			broth & wine & fish & bread & porpoise\\
			onion & sage & bone & vinegar & almond\\
			saffron & parsley & date & blood & milk\\
			 & hyssop & cod & pepper &\\
			 & pepper & almond & ginger &\\
			 & clove & milk & &\\
			 & cinnamon & sugar & &\\
			 & blood & maces & &\\
			 & & flour & &\\
			 & & rice & &\\
			 & & saffron & &\\
			 & & sandalwood & &\\
			 & & ginger & &\\
			\hline
		\end{tabular}
	\end{center}
	\caption{Five examples from our corpus of 4133 medieval recipes: (a) drawen benes, (b) roo in sene, (c) eles in brasill, (d) sause neyger for maudelard roasted, and (e) furmente with purpeys.}
	\label{table:ex_recipes}
\end{table}
\begin{figure}[t]
	\begin{center}
		\includegraphics[width=0.47\textwidth]{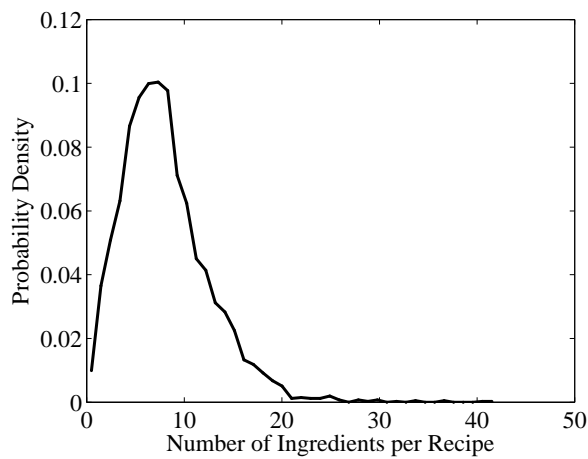}
	\end{center}
	\caption{Probability density function of the number of ingredients per recipe in our medieval corpus.}
	\label{fig:rec_ingred_dist}
\end{figure}
\begin{figure}[t]
	\begin{center}
		\includegraphics[width=0.47\textwidth]{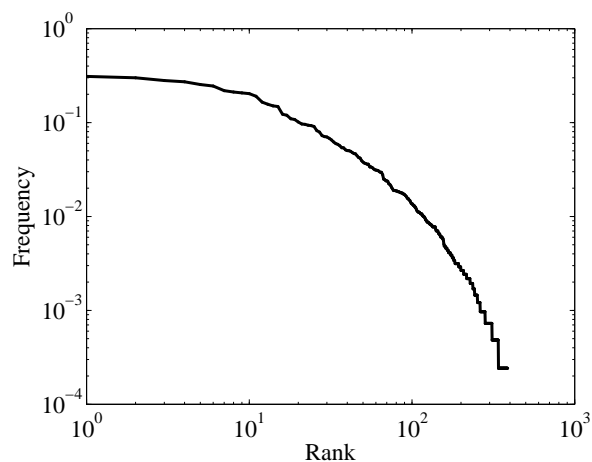}
	\end{center}
	\caption{Rank frequency plot of ingredients in our medieval corpus.}
	\label{fig:rec_ingred_rankfreq}
\end{figure}

\subsection{Medieval Recipe Corpus Creation}
\label{sec:empirical:corpus}

The first step in investigating the flavors of Medieval Europe was to compile a collection of recipes from that age.  In particular, we found recipes in twenty-five different source texts from England, France, Germany, and Italy, from the years 1300 to 1615.  These cookbooks are listed in Table~\ref{table:medieval_cookbooks}.  Next, concordances were generated from the source texts.  From these word lists, ingredient terms were identified, and the remaining parts of speech were discarded.
\begin{figure}[t]
	\begin{center}\scriptsize{
\begin{tabular}{c}
		\begin{tabular}{|l|} \hline
{\bf{2-phenylethanol (=phenethyl alcohol)}}\\
{\bf{safranal (=2,6,6-trimethyl-1,3-cyclohexadienecarbaldehyde)}}\\
{\bf{3,5,5-trimethyl-2-cyclohexen-1-one (=isophorone)}}\\
{\bf{hexadecanoic acid (=palmitic acid)}}\\
{\bf{2,6,6-trimethyl-2-cyclohexene-1,4-dione}}\\
{\bf{(Z,Z)-9,12-octadecadienoic acid (=linoleic acid)}}\\
{\bf{(Z,Z,Z)-9,12,15-octadecatrienoic acid (=linolenic acid)}}\\
naphthalene\\
2,4,6-trimethylbenzaldehyde (=mesitylaldehyde)\\
2,6,6-trimethyl-1,4-cyclohexadienecarbaldehyde\\
6,6-dimethyl-2-methylene-3-cyclohexenecarbaldehyde\\
4-hydroxy-2,6,6-trimethyl-1-cyclohexenecarbaldehyde (=4-hydroxysafranal)\\
3,5,5-trimethyl-3-cyclohexen-1-one\\
3,3,4,5-tetramethylcyclohexanone\\
3,5,5-trimethyl-4-methylene-2-cyclohexen-1-one\\
4-hydroxy-3,5,5-trimethyl-2-cyclohexen-1-one\\
2,3-epoxy-4-(hydroxymethylene)-3,5,5-trimethylcyclohexanone\\
5,5-dimethyl-2-cyclohexene-1,4-dione\\
2,2,6-trimethylcyclohexane-1,4-dione (=3,5,5-trimethyl-cyclohexane-1,4-dione)\\
2-hydroxy-3,5,5-trimethyl-2-cyclohexene-1,4-dione\\
2-hydroxy-4,4,6-trimethyl-2,5-cyclohexadien-1-one\\
2,6,6-trimethyl-3-oxo-1,4-cyclohexadienecarbaldehyde\\
4-hydroxy-2,6,6-trimethyl-3-oxo-1,4-cyclohexadienecarbaldehyde\\
4-hydroxy-2,6,6-trimethyl-3-oxo-1-cyclohexenecarbaldehyde\\
3-hydroxy-2,6,6-trimethyl-4-oxo-2-cyclohexenecarbaldehyde\\
4-(2,2,6-trimethyl-1-cyclohexyl)-3-buten-2-one\\
4-(2,6,6-trimethyl-1-cyclohexen-1-yl)-3-buten-2-one (=$\beta$-ionone)\\
verbenone (=2-pinen-4-one)\\
octadecanoic acid (=stearic acid)\\
(Z)-9-octadecenoic acid (=oleic acid)\\
2(5H)-furanone (=crotonolactone, 2-buten-4-olide, 4-hydroxy-2-butenoic acid lactone)\\
			\hline
		\end{tabular}\\
(a)\\
\\
		\begin{tabular}{|l|} \hline
{\bf{phenethyl alcohol}}\\
{\bf{2,6,6-trimethylcyclohexa-1,3-dienyl methanal}}\\
{\bf{isophorone}}\\
{\bf{palmitic acid}}\\
{\bf{2,6,6-trimethylcyclohex-2-ene-1,4-dione}}\\
{\bf{9,12-octadecadienoic acid (48\%) plus 9,12,15-octadeca- trienoinc acid (52\%)}}\\
			\hline
		\end{tabular}\\
(b)
\end{tabular}
	}
	\end{center}
	\caption{Flavor compounds in saffron (Crocus sativus L.) from (a) VCF data set and (b) Fenaroli data set.  The compounds in bold appear in both lists.}
	\label{fig:ex_compounds1}
\end{figure}

The terms were then manually placed into one of 391 equivalence groupings based upon plurality (e.g.~`cheese' and `cheeses'), synonyms (e.g.~`mallard' and `duck'), spelling variations (e.g.~`chicken' and `chekin'), and foreign loan words (e.g.~`eyren' and `eggs').  These last two types of grouping were made necessary by the inclusion of source texts written in Middle English.  To build the ingredients lists, each source text was split into individual recipes.  The recipes were compared against the table of equivalence groupings, with words not in the table being discarded. Found words were replaced with a lemmatized equivalent for consistency, with duplicates within a recipe being removed.  Several examples of medieval recipes are given in Table~\ref{table:ex_recipes}. Upon visual inspection, the sets of ingredients are quite different than what one experiences today.  The recipe ingredient preparation procedure was done as conscientiously as possible, but is not without error.

In total, the medieval cookbooks contained 4,133 recipes.  After the text processing and word discardal, 41 recipes in our corpus are rendered blank.  The corpus contains 386 different ingredients ranging from acorn to zedoary.  The distribution of the number of ingredients per recipe is shown in Fig.~\ref{fig:rec_ingred_dist}. The mean is 7.74 ingredients per recipe, the maximum is 42, and the standard deviation is 4.60.  A rank frequency plot of the ingredients in the recipe corpus is given in Fig.~\ref{fig:rec_ingred_rankfreq}.

Medieval recipes were chosen for several reasons.  One reason is that they have much historical interest.  In fact, one of the chefs that we interviewed, Michael Laiskonis, specifically mentioned reading historical cookbooks as inspiration for new dishes.  Another reason is that the medieval primary sources are in the public domain.  In contrast, the recipe sources in \cite{AhnABB2011}, e.g.~allrecipes.com, are proprietary and data extracted from them cannot be released publically.  

\subsection{Flavor Compound Data}
\label{sec:empirical:compound}

As we have discussed, we are interested in examining the effect of data sets with different properties, and thus we conduct empirical studies with two different flavor compound databases: VCF and Fenaroli.  The previous work on the flavor pairing hypothesis used only Fenaroli \cite{AhnABB2011}.  The first iteration of VCF, the Lists of Volatiles, was compiled by Weurman in 1963 at the Central Institute for Food Research, which is part of the Nederlandse Organisatie voor Toegepast Natuurwetenschappelijk Onderzoek (TNO).  It is continually updated and enhanced by analytical chemists at TNO.  We use version 14.1 which was released in January 2013.  We scraped and parsed the flavor compound data from the online repository http://www.vcf-online.nl.  

\begin{figure}[t]
	\begin{center}
		\includegraphics[width=0.47\textwidth]{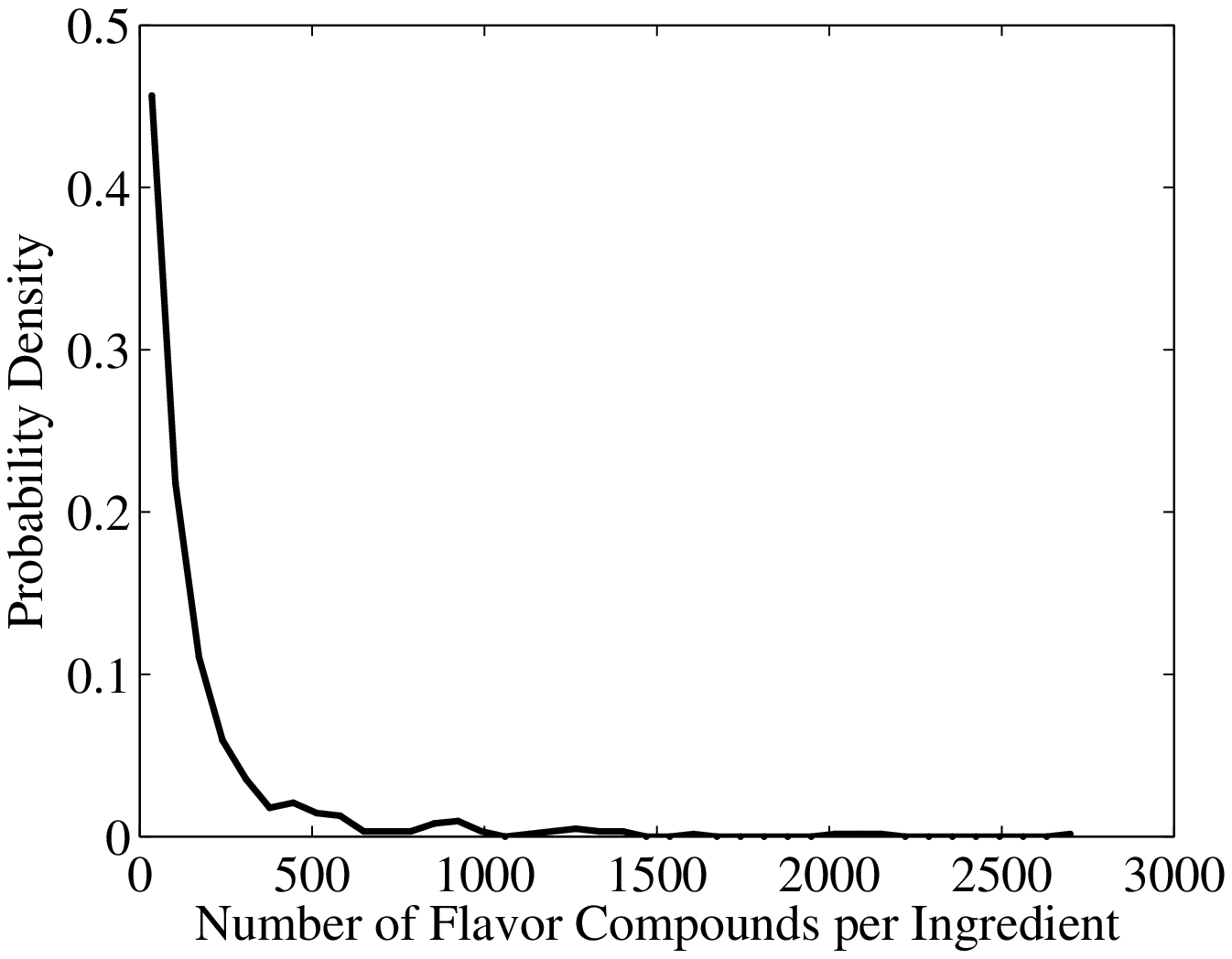}
	\end{center}
	\caption{Probability density function of the number of compounds per ingredient in VCF.}
	\label{fig:vcf_numflavcomp_dist}
\end{figure}

\begin{figure}[t]
	\begin{center}
		\includegraphics[width=0.47\textwidth]{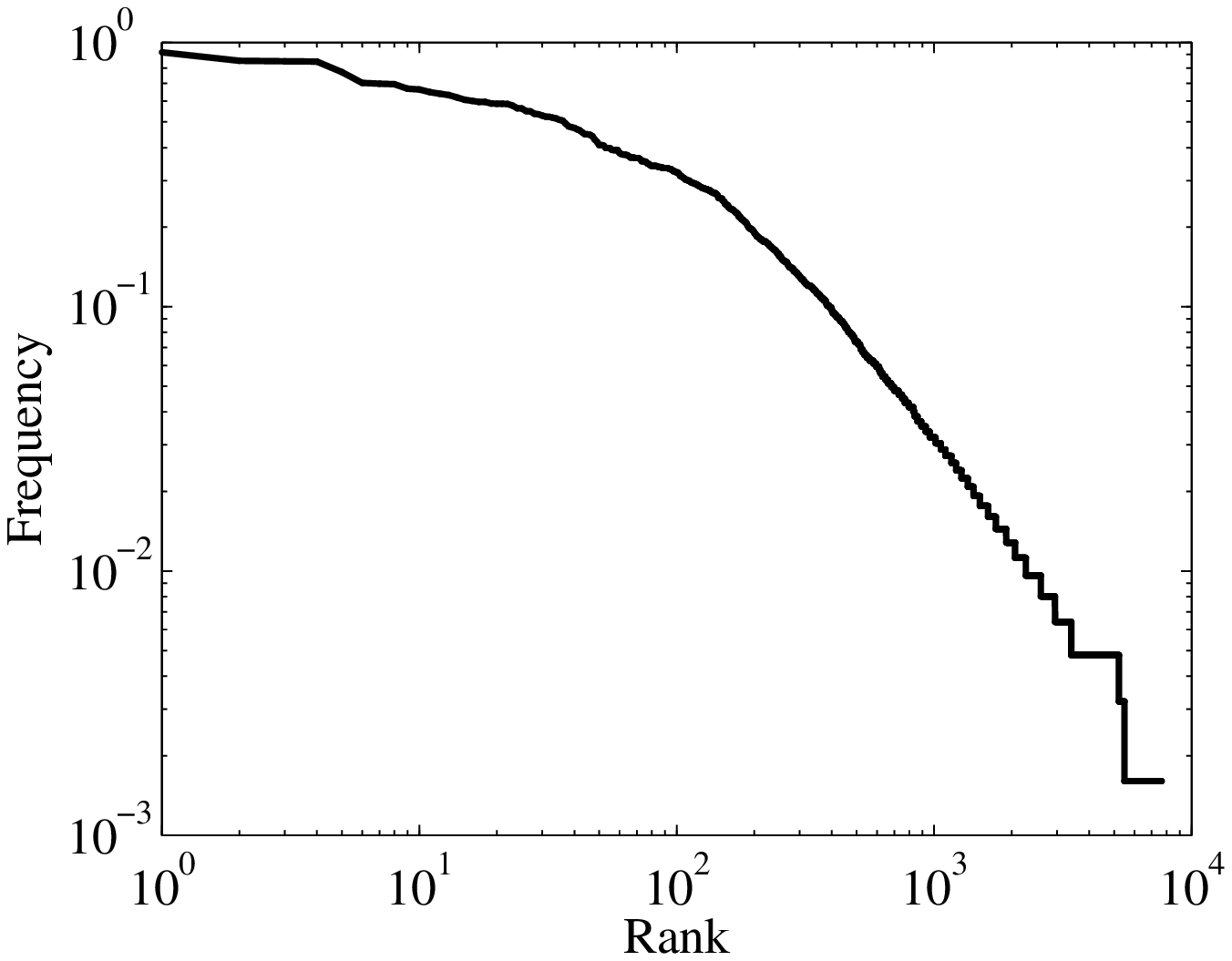}
	\end{center}
	\caption{Rank frequency plot of compounds in VCF.}
	\label{fig:vcf_numflavcomp_rankfreq}
\end{figure}

VCF contains 522 food products and 102 food product categories, which we take together as 624 ingredients.  It also contains 7,647 unique flavor compounds.  Examples of flavor compound listings per ingredient are given in Fig.~\ref{fig:ex_compounds1}(a) and Fig.~\ref{fig:ex_compounds2}(a) for saffron and almond respectively.
\begin{figure}[t]
	\begin{center}\scriptsize{
\begin{tabular}{c}
		\begin{tabular}{|l|} \hline
{\bf{$\alpha$-ionone}}\\
{\bf{2-acetylpyrrole (=methyl 2-pyrrolyl ketone)}}\\
{\bf{phenol (=hydroxybenzene)}}\\
{\bf{furfuryl alcohol (=(2-furyl)-methanol, 2-furanmethanol)}}\\
{\bf{methyl 2-furancarboxylate}}\\
{\bf{furfuryl acetate (=2-furanmethanol acetate)}}\\
{\bf{6,7-dihydro-5-methyl-5H-cyclopentapyrazine}}\\
{\bf{3-methyl-1,2-cyclopentanedione (=cyclotene)}}\\
{\bf{trimethylpyrazine}}\\
hexane\\
benzaldehyde\\
4-hydroxy-4-methyl-2-pentanone (=diacetone alcohol)\\
(E)-$\beta$-ionone\\
2-pyrrolecarbaldehyde (=2-formylpyrrole)\\
(2-furyl)pyrazine\\
2,5-dimethylpyrazine\\
2,6-dimethylpyrazine\\
2-(2-furyl)-3-methylpyrazine\\
furfural (=2-formylfuran, 2-furancarbaldehyde, 2-furaldehyde)\\
5-(hydroxymethyl)furfural\\
4-hydroxy-2-(hydroxymethyl)-5-methyl-3(2H)-furanone\\
2-acetylfuran (=2-furyl methyl ketone, 1-(2-furyl)ethanone)\\
			\hline
		\end{tabular}\\
(a)\\
\\
		\begin{tabular}{|l|} \hline
{\bf{a-ionone}}\\
{\bf{methyl-2-pyrrolyl ketone}}\\
{\bf{phenol}}\\
{\bf{furfuryl alcohol}}\\
{\bf{methyl furoate}}\\
{\bf{furfuryl acetate}}\\
{\bf{5h-5-methyl-6,7-dihydrocyclopenta(b)pyrazine}}\\
{\bf{methylcyclopentenolone}}\\
{\bf{2,3,5-trimethylpyrazine}}\\
acetylpyrazine\\
1-tyrosine\\
b-ionone\\
l-histidine\\
			\hline
		\end{tabular}\\
(b)
\end{tabular}
	}
	\end{center}
	\caption{Flavor compounds in almond (roasted) (Prunus amygdalus) from (a) VCF data set and (b) Fenaroli data set.  The compounds in bold appear in both lists.}
	\label{fig:ex_compounds2}
\end{figure}
The distribution of the number of flavor compounds per ingredient is shown in Fig.~\ref{fig:vcf_numflavcomp_dist}.
The minimum number is 1 (lobster), the maximum is 2,733 (wine), the median is 83, the mean is 175.95, and the standard deviation is 285.93.  A rank frequency plot of compounds in VCF is given in Fig.~\ref{fig:vcf_numflavcomp_rankfreq}.

\begin{figure}[t]
	\begin{center}
		\includegraphics[width=0.47\textwidth]{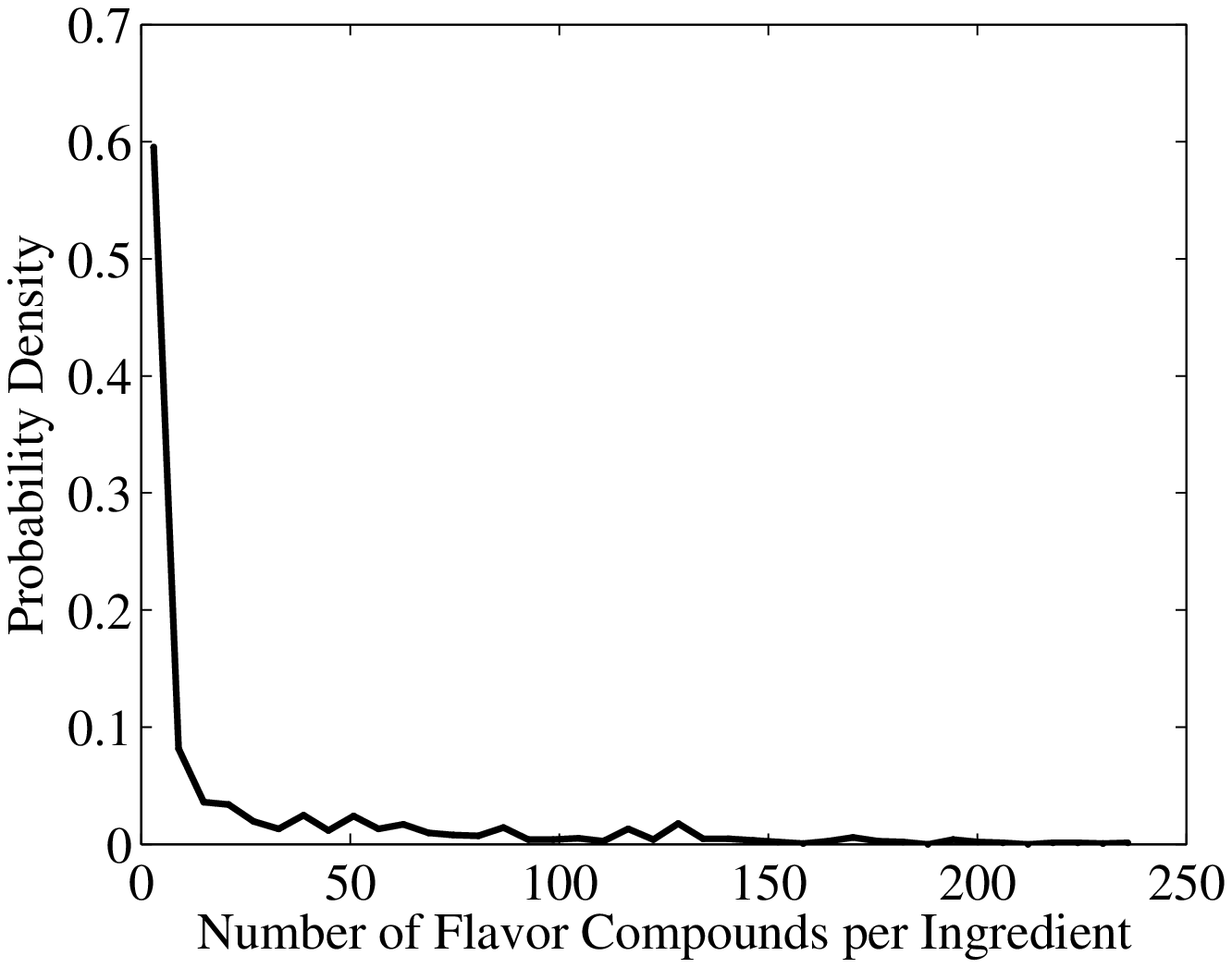}
	\end{center}
	\caption{Probability density function of the number of compounds per ingredient in Fenaroli.}
	\label{fig:fen_numflavcomp_dist}
\end{figure}

\begin{figure}[t]
	\begin{center}
		\includegraphics[width=0.47\textwidth]{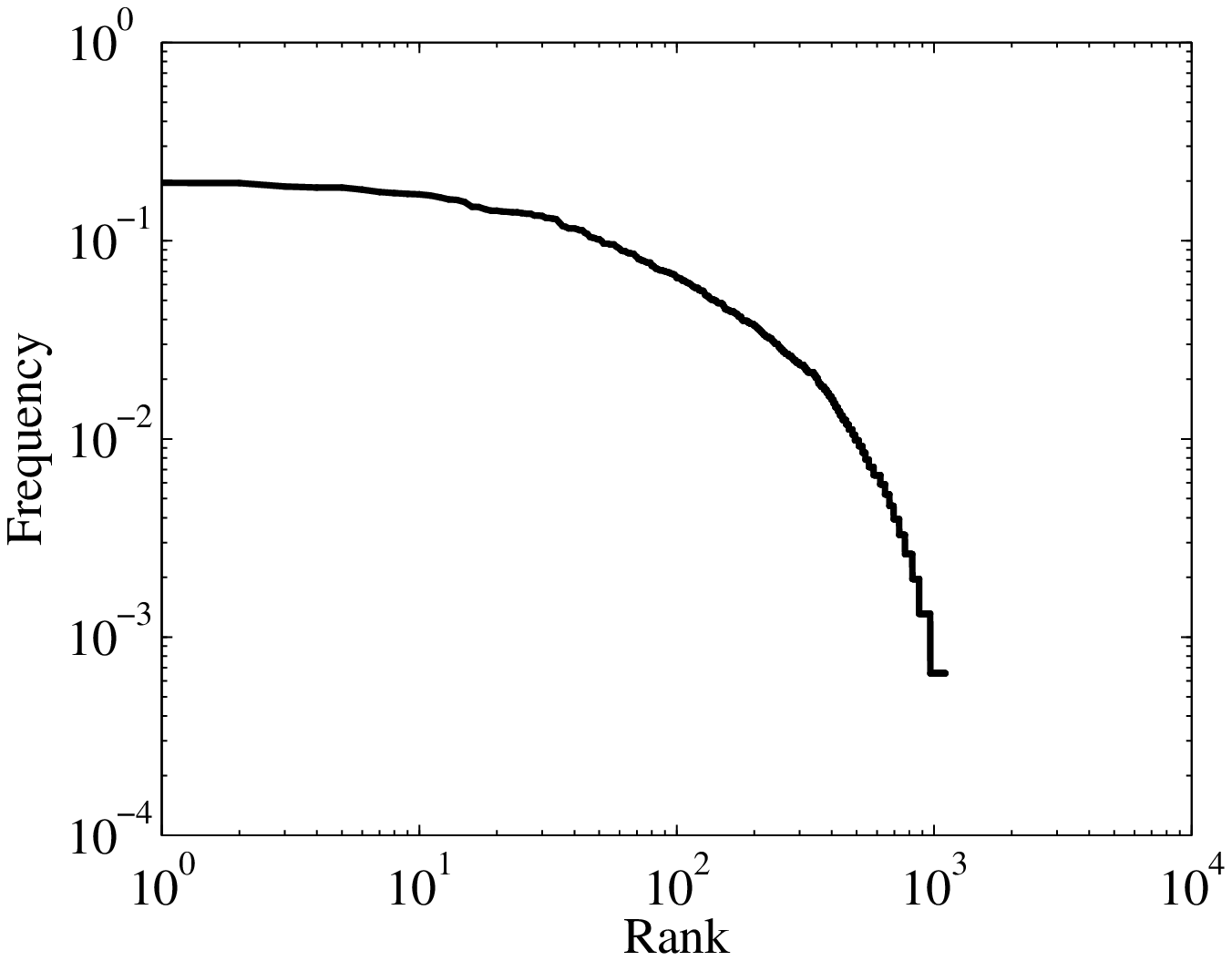}
	\end{center}
	\caption{Rank frequency plot of compounds in Fenaroli.}
	\label{fig:fen_numflavcomp_rankfreq}
\end{figure}

The other flavor compound source is the fifth edition of Fenaroli's Handbook of Flavor Ingredients \cite{Burdock2004} as processed by \cite{AhnABB2011}.  The first edition of this work was published in 1971 and there also now exists a sixth edition.  This data set has 1,530 ingredients and 1,107 flavor compounds.  The distribution of the number of compounds per ingredient is shown in Fig.~\ref{fig:fen_numflavcomp_dist}.
The maximum number of flavor compounds per ingredient is 239 (black tea), the median is 2, the mean is 24.04, and the standard deviation is 43.07.  The rank frequency plot is given in Fig.~\ref{fig:fen_numflavcomp_rankfreq}.
The compound lists for saffron and almond from Fenaroli are given in Fig.~\ref{fig:ex_compounds1}(b) and Fig.~\ref{fig:ex_compounds2}(b).

Fenaroli has a greater number of ingredients than VCF but a much smaller number of flavor compounds.  The average number of compounds per ingredient detected by Fenaroli is also much less than VCF.  The mean to standard deviation ratio for both sets is similar and the shape of the distributions and rank frequency plots is also similar.  The real key difference is in the coverage of the data sets reflected in the axis labels of the plots.  Looking at the two example ingredients, saffron and roasted almond, we see that all compounds listed for saffron in Fenaroli also appear in VCF,\footnote{The names of compounds may not match exactly, but are matched chemically using the Chemical Abstracts Service registry.  The orders of the matching molecules correspond in the tables.} but there are additional compounds in VCF.  Similarly, most Fenaroli compounds for roasted almond appear in VCF whereas VCF has a greater number that do not appear in Fenaroli.  We examine the effect of such a differences in data on quantification of flavor pairing in Section \ref{sec:empirical:pairing}.

\subsection{Ingredient Matching}
\label{sec:empirical:matching}

The final piece of data preparation is matching the names of ingredients from the medieval recipes and the two chemical compound data sets.  For Fenaroli, we used the ingredient names of \cite{AhnABB2011} and did a simple string match to the ingredient names in our medieval corpus.  We were able to match 157 ingredients and were unable to match 229 ingredients.  We note that Ahn et al.~associate the compounds in essential oils and extracts to the original ingredient and include the flavor compounds of more general ingredients into more specific ingredients.  For VCF, we did manual processing of the ingredients and were able to match 191 ingredients.  In VCF, ingredients are sometimes part of larger ingredient categories; following a similar philosophy as Ahn et al., we matched to categories when possible.  One major factor in the higher VCF match rate is that the medieval corpus contains forty-two different fish, e.g.~anchovy, bass, dace, hake, and whiting.  We matched all of these fish to the VCF product category fish.  Although the overall database of Fenaroli has more ingredients than VCF, the difference is no longer significant after matching to the medieval corpus.

\subsection{Flavor Pairing Analysis}
\label{sec:empirical:pairing}

\begin{figure}[t]
	\begin{center}
		\includegraphics[width=0.47\textwidth]{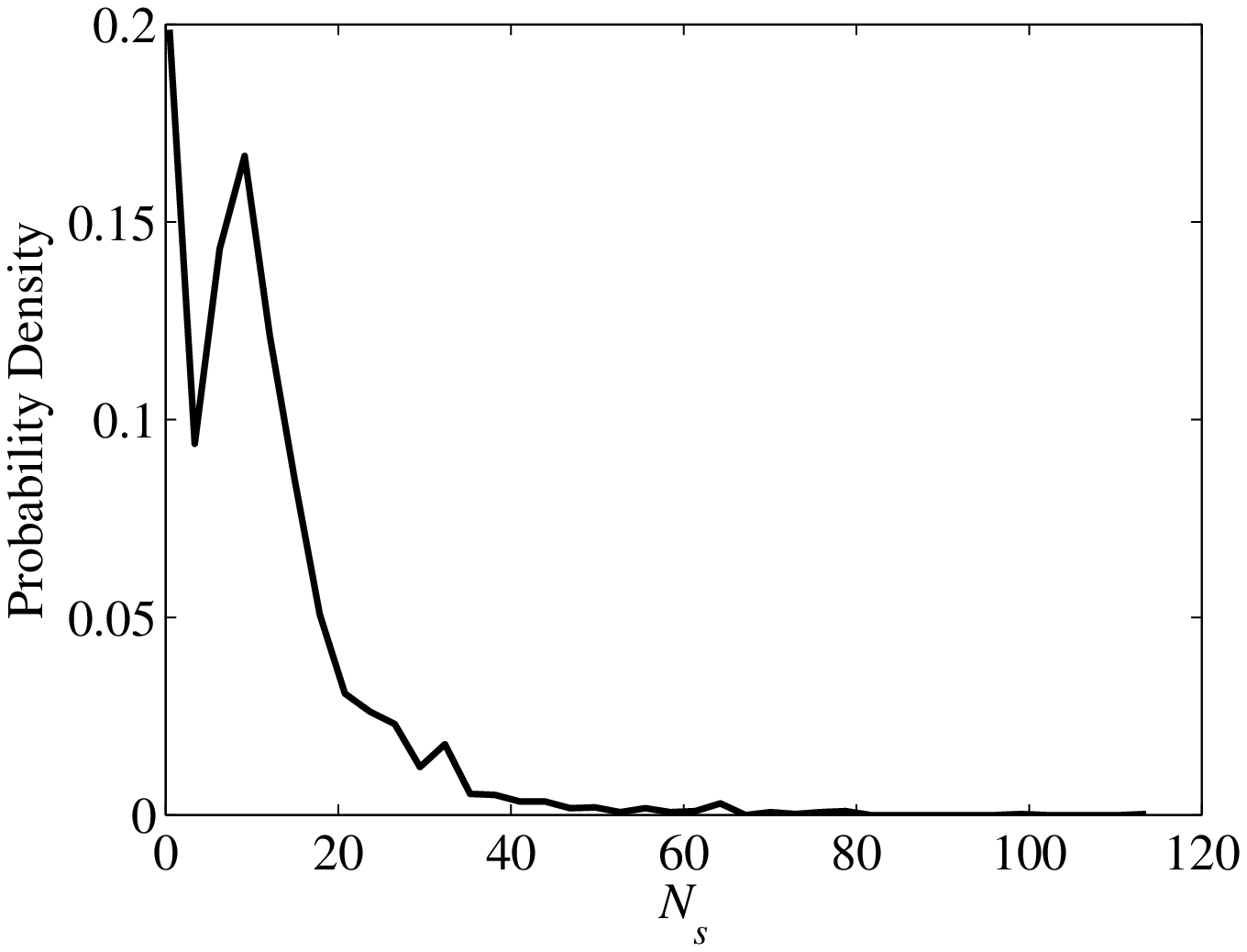}
	\end{center}
	\caption{Probability density function of the number of average shared flavor compounds per recipe from the Fenaroli data set.  The mean $\bar{N}_s = 11.26$.}
	\label{fig:fen_Ns_dist}
\end{figure}

\begin{figure}[t]
	\begin{center}
		\includegraphics[width=0.47\textwidth]{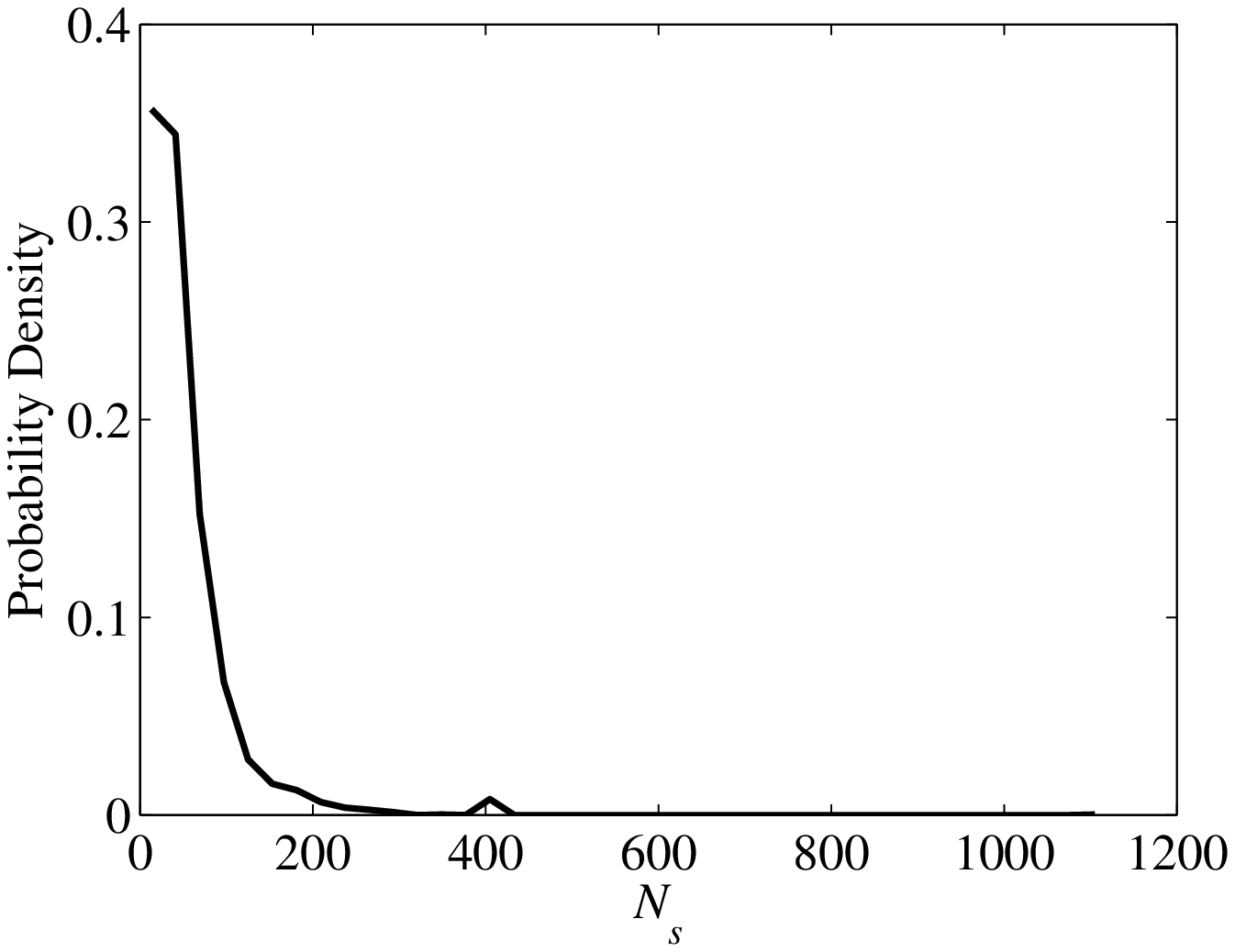}
	\end{center}
	\caption{Probability density function of the number of average shared flavor compounds per recipe from the VCF data set.  The mean $\bar{N}_s = 51.42$.}
	\label{fig:vcf_Ns_dist}
\end{figure}

With all data collected, prepared, and matched, we can perform the statistical analysis described in Section~\ref{sec:pairing:math}.  We first calculate the average number of shared compounds among the over four thousand recipes in our medieval corpus.  The distribution of $N_s(R)$ using the Fenaroli data is shown in Fig.~\ref{fig:fen_Ns_dist}.
The distribution using the VCF data is shown in Fig.~\ref{fig:vcf_Ns_dist}.
The average across the corpus is calculated as $\bar{N}_s^{\rm{real}} = 11.26$ for Fenaroli and $\bar{N}_s^{\rm{real}} = 51.42$ for VCF.  

\begin{figure}[ht]
	\begin{center}
		\includegraphics[width=0.47\textwidth]{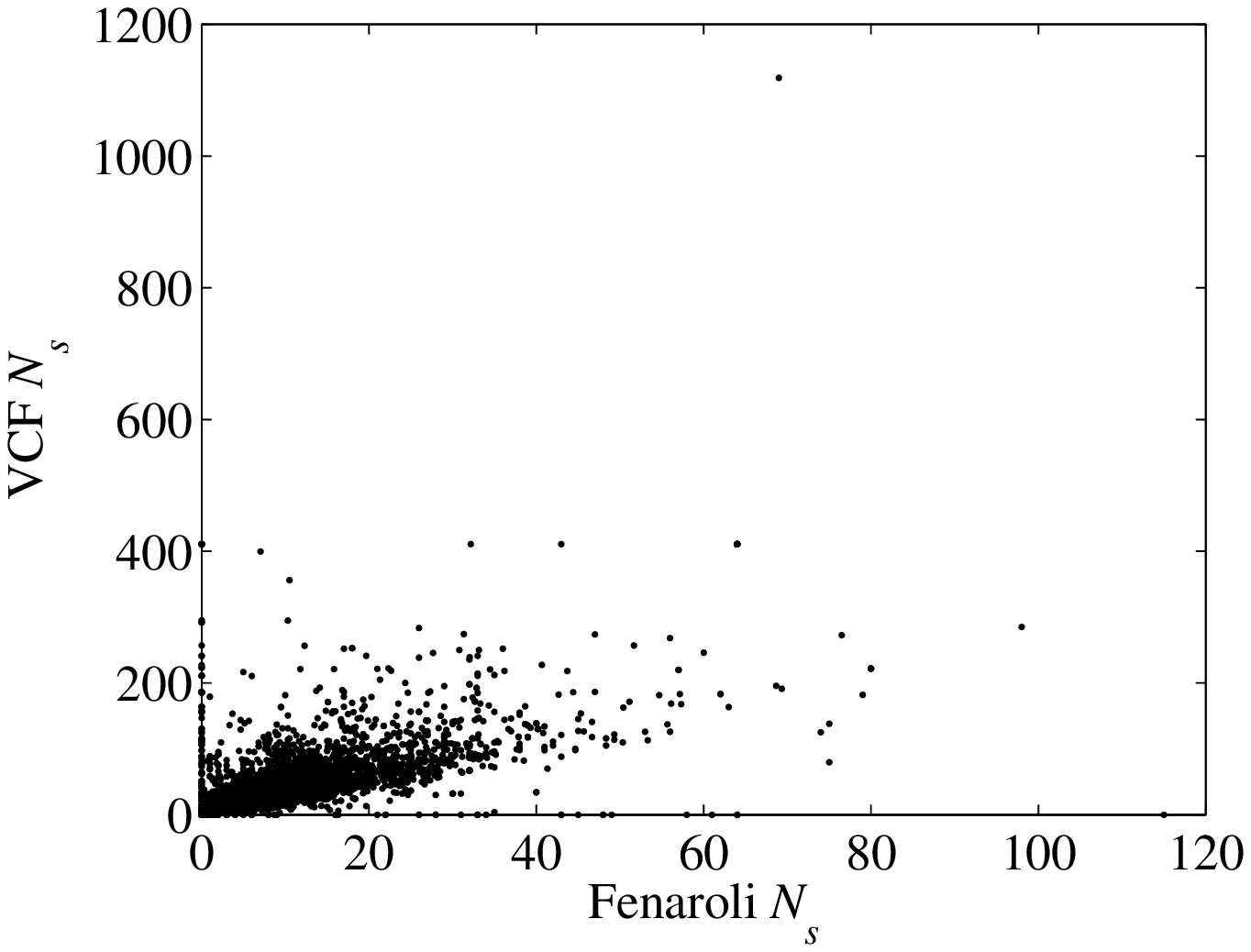}
	\end{center}
	\caption{Scatter plot of the number of average shared flavor compounds per recipe calculated using the Fenaroli and VCF data sets.}
	\label{fig:fenvcf_Ns_scatter}
\end{figure}

The values are quite different due to VCF containing so many more flavor compounds.  We can examine how correlated the $N_s(R)$ values are when using the two different flavor compound data sets.  A scatter plot is shown in Fig.~\ref{fig:fenvcf_Ns_scatter}.
The correlation coefficient between the two is $0.542$.  Since most of the $N_s(R)$ values are small, we can also examine the correlation after a logarithmic transformation.  The correlation coefficient between the $\log(N_s(R)+1)$ values using the two data sets is $0.688$.  The shared compound calculation using the two different flavor compound data sets yields similar results, but not a full correlation.

Now to understand the meaning of the $\bar{N}_s^{\rm{real}}$ values, we must also calculate $\bar{N}_s^{\rm{rand}}$ for the two compound sets.  Using the exact same instantiation of random medieval ingredient sets, we find $\bar{N}_s^{\rm{rand}} = 4.54$ for Fenaroli and $\bar{N}_s^{\rm{rand}} = 54.17$ for VCF, yielding $\Delta N_s = 6.72$ for Fenaroli and $\Delta N_s = -2.75$ for VCF. 

These delta values lead to opposite conclusions.  On one hand, using the Fenaroli data, we see a very strong positive indicator of the flavor pairing hypothesis in Medieval Europe.  A value of 6.72 is much larger than for any modern cuisine reported in \cite{AhnABB2011}.  (The $\bar{N}_s^{\rm{real}}$ value for Medieval Europe is actually quite similar to modern North America, but the $\bar{N}_s^{\rm{rand}}$ value for Medieval Europe is smaller.)  On the other hand, using the VCF data, we obtain a negative $\Delta N_s$, which means that ingredients that don't share many flavor compounds are used together.  

We can also calculate and examine the individual ingredient contributions.  Table~\ref{table:contributors} list the top and bottom fifteen contributors using the two data sets.
\begin{table}
	\begin{center}\small
		\begin{tabular}{|l|c|c|c|c|} \hline
			& Top Fen & Top VCF & Bottom Fen & Bottom VCF\\\hline
			1 & whale & halibut & filbert & valerian \\\hline
			2 & blackberry & dace & lentil & buttermilk \\\hline
			3 & bacon & thorneback & octopus & horseradish \\\hline
			4 & haddock & sole & valerian & eggplant \\\hline
			5 & tuber & hake & horseradish & oregano \\\hline
			6 & beer & turbot & caviar & chicory \\\hline
			7 & salmon & mullet & oregano & cuttlefish \\\hline
			8 & cider & carp & chickpea & caviar \\\hline
			9 & beef & dogfish & cuttlefish & clam \\\hline
			10 & strawberry & ray & vervain & lentil \\\hline
			11 & cod & shad & nettle & barley \\\hline
			12 & herring & trout & buttermilk & turkey \\\hline
			13 & cheese & citron & clam & minnow \\\hline
			14 & grape & gurnard & pennyroyal & prawn \\\hline
			15 & bean & bream & rue & scallop \\\hline
		\end{tabular}
	\end{center}
	\caption{Top and bottom fifteen contributing ingredients to medieval cuisine.}
	\label{table:contributors}
\end{table}
The bottom fifteen contributors are fairly stable with respect to the two chemical databases, but the top fifteen contributors are different.  The VCF list is dominated by fish, whereas the Fenaroli list does have many fish, but other things as well.  Most of the fish at the top of the VCF list were not matched using Fenaroli data.  This difference may be the main contributor for the conflicting results.

\section{Conclusion}
\label{sec:conclusion}

In this work, we have examined food ingredients that appear in medieval recipes, focusing on how many chemical flavor compounds the ingredients share.  Our contribution is studying the reasons and effects of dirty data, in particular finding that conclusions can be reversed by differences in data quality.  Specifically, we have tested the hypothesis that food ingredients that share many flavor compounds go together in dishes.  

Using a sparser and more incomplete chemical database and matching procedure, we find the hypothesis to be true in Medieval Europe.  Moreover, in comparing with analysis of modern regional cuisines using the same exact chemical database, we find the pairing to be stronger in the medieval period than in modern times.  The main difference is not in the level of pairing in the recipes, but in the lack of potential pairing in the available ingredients as expressed through random recipe ensembles.  As is known historically, the number of ingredients available after the Columbian Exchange, including in the modern world, is much greater than before.  The results we obtain bear this fact out.  Even though medieval cooks had a more difficult job because there were fewer paired ingredients, they were able to achieve the same level of flavor compound pairing.  After the exchange and the introduction of a boatload of new ingredients, Western cooks have maintained the pairing level, but increased variety. We can conjecture that there is some combination of pairing and variety or balance that chefs aim to achieve; by having more ingredients, they are able to more easily satisfy the pairing and turn their attention to variety and balance.
 
In future work, it would be interesting to see whether this inference, comparison, and conjecture holds when analyzing modern recipes using the more complete chemical database we have utilized to study medieval cuisine.  We have seen here that the quality of the raw data and quality of the data preparation have a fundamental downstream effect on analysis.  The more complete dataset has indicated the opposite: that ingredients with shared compounds are not over-represented in recipes.  Understanding this result requires more detailed study.

Scientifically validated design principles are important for using computational techniques in generating flavorful, novel, and healthy culinary recipes.  These design principles, however, are often derived from large-scale data analysis, and so there is a need for complete and accurate data sources.  In this work on medieval cuisine, we found different results using different flavor compound databases.  From the point of view of computational creativity for culinary recipes, if we want to generate foods that people from Medieval Europe might find flavorful, this leaves us in a bit of a dilemma.  Should we be promoting flavor pairing or not.  

Broadly speaking, computational creativity algorithms have two phases: first generating combinatorially many new ideas, and then evaluating the ideas on metrics of quality and novelty.  Each domain of creativity, whether music, literature, or food recipes, needs a defined notion of quality.  Flavor pairing is a putative quality metric for cooking, but our analysis here is not conclusive and indicates that there is more to the story.  

\section*{Acknowledgments}

The authors thank Nizar Lethif for assistance in preparing the VCF data set and Y.-Y.~Ahn, Sebastian Ahnert, Cindi Avila, Debarun Bhattacharjya, James Briscione, Yi-Min Chee, Chris Gesualdi, Michael Laiskonis, Aleksandra Mojsilovi{\'c}, Florian Pinel, Angela Sch{\"o}rgendorfer, and Dan Stone for discussions.

\bibliographystyle{named}
\bibliography{medieval}

\end{document}